# Ultrafast formation of transient 2D diamond-like structure in twisted bilayer graphene

Duan Luo[1,2,3†], Dandan Hui[2,3†], Bin Wen[4], Renkai Li[5]*, Jie Yang[5], Xiaozhe Shen[5], Alexander Hume Reid[5], Stephen Weathersby[5], Michael E. Kozina[5], Suji Park[5], Yang Ren[6], Troy D. Loeffler[1], S. K. R. S. Sankaranarayanan[1], Maria K. Y. Chan[1], Xing Wang[2], Jinshou Tian[2,7], Ilke Arslan[1], Xijie Wang[5], Tijana Rajh[1]*, Jianguo Wen[1]*

1. Center for Nanoscale Materials, Argonne National Laboratory, Lemont, IL 60439, USA.

2. Key Laboratory of Ultra-fast Photoelectric Diagnostics Technology, Xi'an Institute of Optics and Precision Mechanics, Chinese Academy of Sciences, Xi'an 710119, China.

3. University of Chinese Academy of Sciences, Beijing 100049, China.

4. State Key Laboratory of Metastable Materials Science and Technology, Yanshan University, Qinhuangdao 066004, China.

5. SLAC National Accelerator Laboratory, Menlo Park, CA 94025, USA.

6. X-ray Science Division, Argonne National Laboratory, Lemont, IL 60439, USA.

7. Collaborative Innovation Center of Extreme Optics, Shanxi University, Taiyuan 030006, China.

*Corresponding author. Email: lrk@slac.stanford.edu(R.K.L.); rajh@anl.gov(T.R.); jwen@anl.gov(J.G.W.)

†These authors contributed equally to this work.

Dec 1, 2021

**Due to the absence of matching carbon atoms at honeycomb centers with carbon atoms in adjacent graphene sheets, theorists predicted that a sliding process is needed to form AA, AB', or ABC stacking when directly converting graphite into $sp^3$ bonded diamond. Here, using twisted bilayer graphene (TBG), which naturally provides AA and AB' stacking configurations, we report the ultrafast formation of a transient 2D diamond-like structure (which is not observed in aligned graphene) under femtosecond laser irradiation. This photo-induced phase transition (PIPT) is evidenced by the appearance of new bond lengths of 1.94 Å and 3.14 Å in the time-dependent differential pair distribution function (ΔPDF) using MeV ultrafast electron diffraction (UED). Molecular dynamics and first principles calculation indicate that $sp^3$ bonds nucleate at AA and AB' stacked areas in moiré pattern. This work sheds light on the direct graphite-to-diamond transformation mechanism, which has not been fully understood for more than 60 years.**

Atomically-thin twisted vdW crystals with a rotational misalignment introduce a moiré pattern[1-11], leading to intriguing electronic properties such as unconventional superconductivity[2,3] and correlated insulating states[4] by modulating the band structure of the material. New optical phenomena such as moiré excitons[5-8] in vdW heterostructures were also reported at different twisted angle configurations. Meanwhile, optical control of structural and electronic properties, especially through PIPT, has recently attracted tremendous attention due to its great potential in functionalization of 2D materials[13-16]. A detailed study of the changes in bonding network and the resulting metastable dynamics triggered by photoexcitation is necessary but challenging since it's an intrinsically out-of-equilibrium process[17]. These make the ultrafast probing of the fundamental light-matter interactions in twisted 2D vdW materials, especially TBG, very attractive.

In perfectly aligned graphene structures with AB stacking, the absence of matching carbon atoms at honeycomb centers with carbon atoms in adjacent layers hinders direct transformation into diamond. Therefore, extensive theoretical research suggests that the direct transformation from hexagonal graphite to diamond must undergo a sliding process to AB' stacking[18], AA stacking[19], or ABC stacking[20] configurations, followed by boat or chair buckling (Fig. S1). On the other hand, TBG naturally creates a moiré pattern with periodic metastable AA and AB' stacked local areas (Fig. 1a-1b). Upon excitation, these areas are expected to easily form $sp^3$ rich metastable structures (Fig. 1c). We chose an fs laser to initiate the structural phase transition,[21-23] since it has been demonstrated that fs laser irradiation has the ability to convert graphite to diamond structure at ambient temperature and pressure. Additionally, ultrashort laser excitation can be, coupled with an ultrafast electron pulse in a pump-probe configuration to enable the study of the structural dynamics in real time[24,25].

Here we report the photo-induced $sp^2$-$sp^3$ transformation in TBG where the moiré pattern plays an important role. Through analysis of the ΔPDF, an ultrafast formation of diaphitene from TBG was observed after fs laser irradiation. Similar to its corresponding bulk form (called diaphite[23,26]), diaphitene exhibits interlayer $sp^3$ bonding but differs from a diamond lattice with C-C bond length of 1.54 Å. Furthermore, the recorded dynamics revealed that the transition was completed within 330 fs and the structure was maintained as diaphitene for at least 2 ns and eventually recovered back to the graphene structure within 5.56 ms (pump frequency 180 Hz).

Large area TBG samples were prepared by stacking two chemical vapor deposition (CVD) monolayer graphene sheets on gold grids with a holey carbon support film (Supplementary

Materials). Each graphene sheet in TBG consists of numerous grains with a typical grain size of ~5 μm. Twist angles are determined from 137 different overlapped domains using selected-area electron diffraction (SAED) with a 500 nm aperture to be mainly concentrated at ~12° (Fig. S2). Through analysis of the intensity profile along the green line marked in SAED pattern, we can determine almost all the areas are stacked by two monolayer graphene since the intensity ratios $I_{\{100\}}/I_{\{110\}}>1$[27]. UED experiments were performed using an electron pulse (~120 μm probe size, ~130 fs pulse length) at normal incidence. The UED diffraction pattern of TBG consists of Debye-Scherrer rings (Fig. 2a), since the grain size in TBG (~5 μm) is much smaller than the probe size of electron pulse (~120 μm). The time-dependent intensity of {100} and {110} diffraction rings after photoexcitation at an incident fluence of ~9.7 mJ/cm² is shown in Fig. 2b. Similar phenomena was also observed at a fluence of ~4.2 mJ/cm² (Fig. S3). The intensity changes clearly show an initial ultrafast component followed by a slower one. The time evolution was thus fitted by bi-exponential functions $\frac{\Delta I}{I_0}(t) = A(1 - e^{-t/\tau_1}) + B(1 - e^{-t/\tau_2})$. The extracted time constants are $\tau_1$ =115 fs and $\tau_2$ =585 fs for the {100} peak and $\tau_1$ =188 fs and $\tau_2$ =10.8 ps for the {110} peak (Fig. S3), respectively. Such a rapid intensity rise in the {100} peak and a drop in {110} peak indicate that this process is not a thermal effect, but possibly induced by in-plane buckling during interlayer contraction. The temporal evolution of the magnitude of the momentum transfer vector $Q$ (i.e., $Q = \frac{2\pi}{d_{hkl}}$) is illustrated in Fig. 2c. Both the {100} and {110} peaks shift to a lower $Q$ within ~330 fs. Photo-induced electron-phonon scattering which heats the lattice and results in thermal expansion can also result in an increase of lattice spacing. However, thermal expansion typically happens at a longer timescale. The observed reduction in $Q$ (increase in the in-plane lattice spacing) indicates a possible conversion of graphene sheets into an hexagonal diamond (HD)-like structure (Fig. 2c), since the lattice spacings in HD (2.18 Å (100) and 1.26 Å (110)) is larger than that in graphene sheets (2.13 Å (100) and 1.23 Å (110)). Repeated pump-probe experiments show that the change is reversible, but it does not revert to graphene even after ~2 ns as shown in the long-runtime experiment (Fig. S3). To rule out the possible contribution due to the supporting amorphous carbon (AC) film, we performed ultrafast experiments on the AC film. No obvious changes occurred in time-dependent $Q$ of the AC films compared to the TBG (Fig. S3), thus confirming that the structural transformation is truly from the TBG.

In contrast, such an ultrafast structural transformation has not been reported in aligned bilayer and few-layer graphene before. For clear comparison, we carried out UED experiments on free-standing AB-stacked aligned bilayer and few-layer graphene (Fig. S4) under the same conditions. We found that there are no noticeable changes in time-dependent $Q$ compared to TBG (Fig. S5); i.e., no phase transitions occur in aligned graphene sheets. Instead, the {100} diffraction peaks of aligned graphene sheets exhibit a strong shearing mode vibration with a clear oscillation period of 0.8 ps and a damping time of 15 ps, similar to the results reported by Siwick et al[28]. Even doubling the fluence to ~19 mJ/cm² showed no other changes except a larger amplitude.

MeV electrons enable UED from TBG to satisfy a kinematic approximation similar to the ones used in X-ray and neutron diffraction. Therefore, the time-dependent ΔPDF[29] was calculated from the diffraction patterns to further extract the structural dynamics of TBG following photoexcitation. The diffraction-difference method which employs the subtraction of diffraction signals before and after optical excitation was used to highlight these changes (Supplementary Materials). The experimental ΔPDF (at fluence of ~9.7 mJ/cm²) as a function of the pump-probe time delay between -200 fs and 1000 fs with a step size of 67 fs is shown in Fig. 3a, with blue indicating loss and red indicating gain in the ΔPDF at various atom pair distances compared with unexcited TBG. Ultrafast changes in interatomic distances were observed from the ΔPDF. By choosing the appropriate time delays (from Fig. 3a and Fig. S6), we obtained a series of one-dimensional ΔPDFs, as shown in Fig. 3b. The positive peaks (II, IV) represent a significant increase in the ΔPDF at interatomic distances $r$ of 1.94 Å and 3.14 Å, while the negative peaks (I, III) correspond to a reduction of the ΔPDF at interatomic distances at 1.42 Å and 2.46 Å (Fig. 3b). The ultrafast bond length distribution changes reach a maximum at 330 fs (Fig. S6). Static PDF of graphene sheets studied by X-ray and neutron diffraction[30,31] shows three nearest neighbor pair distances in the honeycomb graphene lattice: 1.42 Å, 2.46 Å, 2.84 Å (Fig. 3c, Fig. S7). When compressing two graphene layers to an interlayer distance $d$, the PDF with pair distances < $d$ has the contribution only from monolayer graphene regardless of twist angles and stacking sequences[30,31]. In other words, the appearance of new bond lengths other than 1.42 Å, 2.46 Å and 2.84 Å in PDF reflects the interlayer distance is reached to the smallest valve of new bond lengths. Therefore, the observation of new bonding distances of 1.94 Å and 3.14 Å indicates the ultrafast formation of interlayer $sp^3$ bonds after the fs laser irradiation (Fig. 3c).

To understand the detailed structural evolution during compression, we carried out molecular dynamics (MD) simulations for TBG with different twist angles (0°, 6°, 12°, 18° and 30°) (Supplementary Materials). All TBGs with twist angles other than zero degree with AB stacking showed similar structural dynamics (Fig. S8). Taking a typical angle of 12° as an example (dominant angle in our experimental work), one can see that after excitation, $sp^3$ bonds labeled with red and orange (warm colors) are preferentially formed in local areas with stacking close to AA and AB', but almost no $sp^3$ bonds are formed (shown in blue) in the area with AB stacking from the top-view snapshots (Fig. 4a and Fig. S8). These simulation results support our experimental observations of ultrafast phase transitions in TBG but not in aligned graphene sheets that is composed entirely of AB stacking configuration. The reversible phase transition occurs within ~270 fs, but it takes a much longer time to recover due to the strong covalent interaction of C-C bonds between two graphene layers as shown in the side-view and perspective-view snapshots (Fig. 4b, Fig. S8). This explains why the structural change occurs within 330 fs and remains the transient state longer than 2 ns.

Using the simulated transient structures, we calculated their ΔPDFs by selecting several representative moments (Fig. S8). These calculated ΔPDFs match well with the experimental ones (Fig. 3b), although there is a slight difference above 4 Å between the experimental and simulated ΔPDF. This difference may be caused by distribution of various twist angles in experimental TBG. From transient TBG structure at the time of ~270 fs with maximum numbers of $sp^3$ bonds, a new transient diaphitene (Fig. 3c) is extracted, which explains the emergence of new pair distances of 1.94 Å, 3.14 Å in ΔPDFs.

To reveal the crucial role that TBG played in $sp^2$-$sp^3$ bond changes, generalized stacking fault energy (GSFE) surfaces for bilayer graphene with AB stacking (Fig. S9) and the energy surface as a function of rotation angles (Fig. S10) are evaluated using first principles calculations under different layer spacings (Supplementary Materials). As shown in Fig. 4c and Fig. S9, when the layer spacing is larger than 2.4 Å, the bilayer graphene with AB stacking (ground state) has the lowest free energy, and, due to the Pauli repulsion, AA stacking has the highest free energy. Bilayer graphene with AB' stacking becomes the ground state under compression when the layer spacing is in the range between 2.4 Å and 2.2 Å. Further compressing layers below ~2.0 Å changes the ground state to the AA stacking, due to the strong covalent effect between overlaying C atoms (formation of $sp^3$ bonds). This calculation indicates that bilayer graphene with AA and AB' stacking is more prone to forming $sp^3$ bonds than AB stacking under compression. In addition, the potential energy surface of TBG as a function of the twisted angles shows no significant angle dependence except for the maximum energy at angles of near zero degree with AB stacking (Fig. S10). We also found some smaller energy surfaces with angles around 1.7°, 2.5°, 3.1°, 13° and 22°.

In conclusion, using ultrafast electron diffraction, we show a diamond-like transient state in TBG under fs laser irradiation. We observe appearance of new bonding distances of 1.94 Å and 3.14 Å within 330 fs in time-dependent difference pair distribution function identifying a new transient 2D diamond-like structure consisting of mixed $sp^2$ and $sp^3$ bonds. Such a photoinduced structural transformation is not observed in aligned graphene sheets. On contrary, only lattice vibration was detected in aligned graphene sheets under fs laser irradiation. Molecular dynamics and first principles calculation indicate that the TBG moiré pattern naturally provides stacking configurations like AA and AB' favorable for the preferential formation of $sp^3$ bonds. This new discovery using TBG offers a promising way to open and precisely tune a bandgap in graphene system, enabling the exploration of innovative applications for the future. This also provides new insights into the understanding of the ultrafast optical response of twisted 2D vdW materials. In addition, this work sheds light on the direct graphite-to-diamond transformation mechanism, which has not been fully understood for more than 60 years.

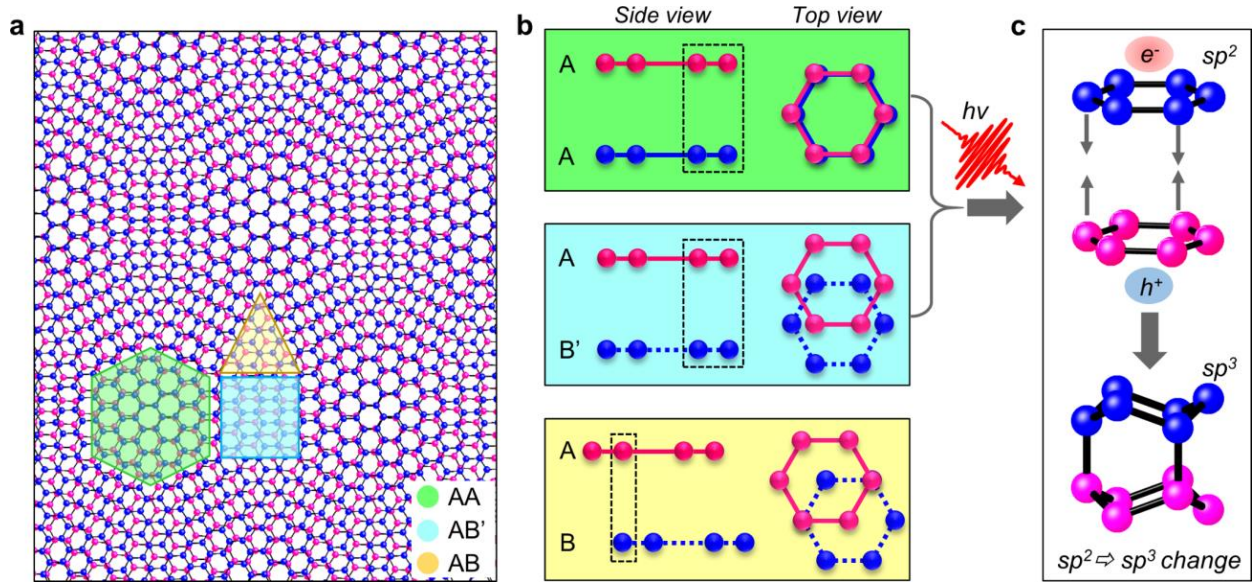

**Fig. 1. Photoinduced structural transition from twisted bilayer graphene (TBG) to 2D diamond structure. a**, Schematic of a TBG moiré pattern, which consists of regions with AB, AA, and AB' stackings as structural details shown in **b**. **c**, Schematic illustration of the photoinduced structural transition in TBG. Subsequent to the fs laser irradiation, interlayer contraction and buckling are induced by generated interlayer charge transfer excitons and then the transformation of $sp^2$ bonds to $sp^3$ bonds occurs. Such a transformation occurs preferentially at regions with AA and AB' stacking rather than at AB stacked areas due to the lack of matching carbon atoms at honeycomb centers between adjacent AB stacked graphene layers (Top view in (**b**)).

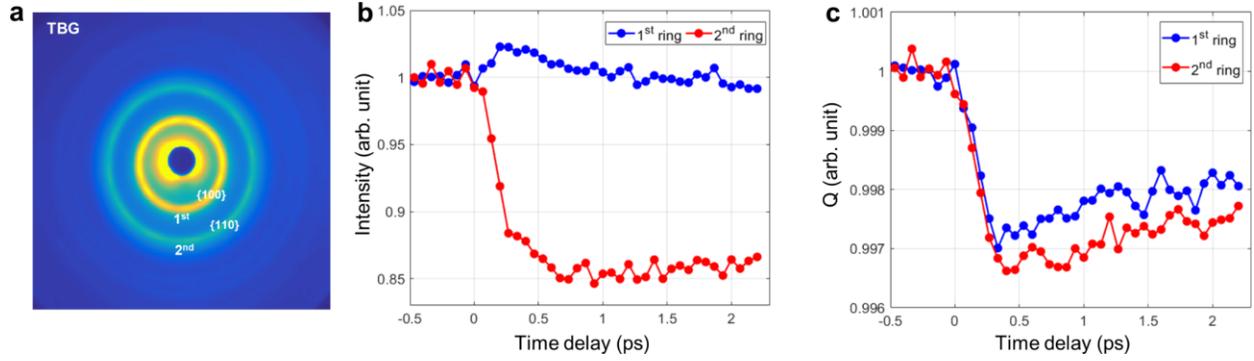

**Fig. 2. Ultrafast structural dynamics in twisted bilayer graphene (TBG). a**, Typical diffraction pattern of TBG observed by MeV UED with first two visible rings. **b**, Diffraction intensity changes as a function of time delays. A rapid intensity rise in the 1$^{st}$ ring {100} and drop in the 2$^{nd}$ ring {110} indicate that this process is not a thermal effect. **c**, Time-dependent $Q$ shift of TBG showing the ultrafast structural transformation within 330 fs.

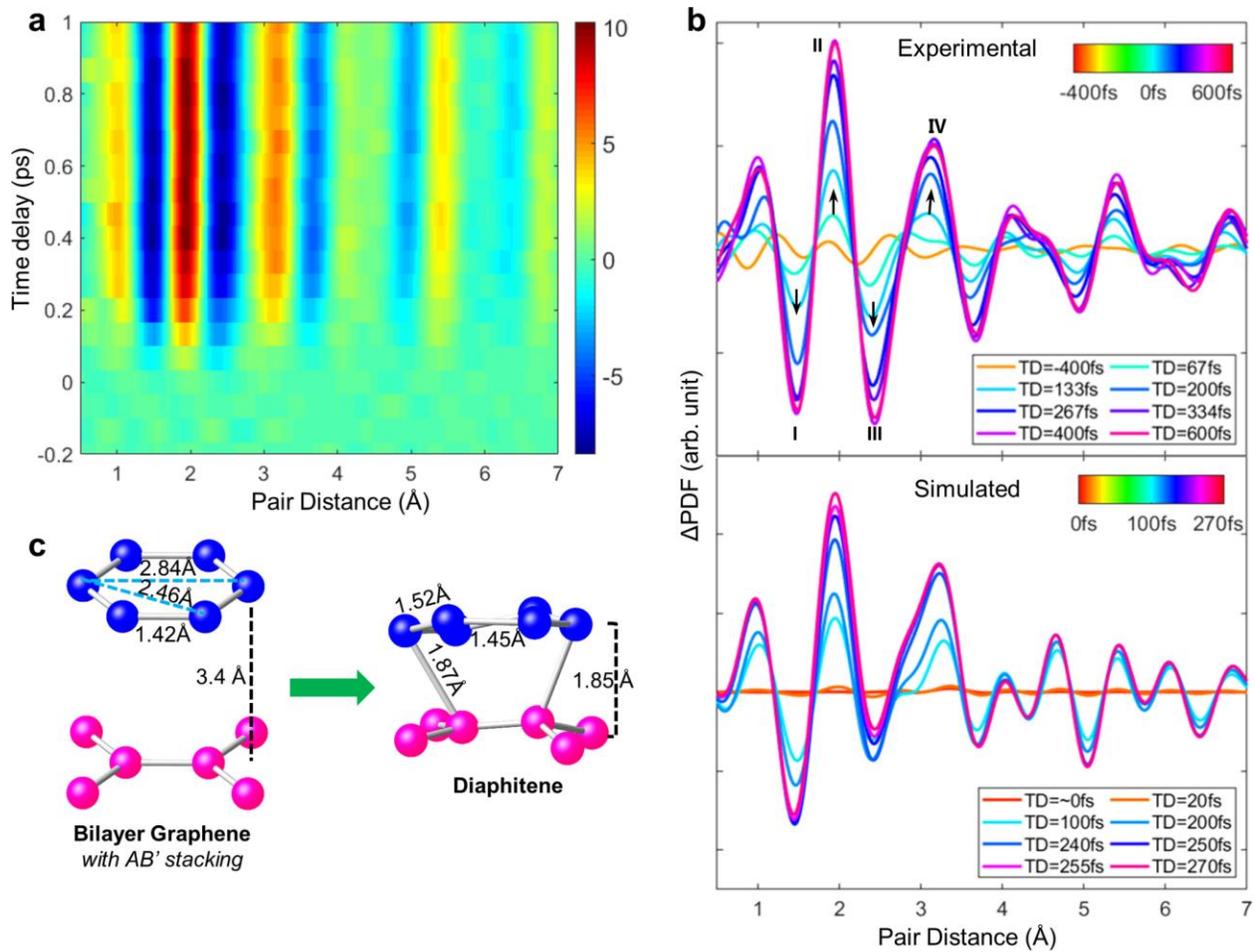

**Fig. 3. Temporal evolution of difference pair distribution functions (ΔPDF). a**, Experimental 2D ΔPDF of twisted bilayer graphene (TBG) showing the intensity increase (warm color) and decrease (cool color) at different interatomic distances and different time delays. **b**, One-dimensional experimental and simulated ΔPDF at different time delays. The observed reduction of 1.42 Å (I) and 2.46 Å (III) bonds in (**b**) matches with the bonds in graphene (**c**) before the transformation. The appearance of 1.94 Å (II) and 3.14 Å (IV) bonds observed in TBG matches with the new diaphitene structure (**c**) extracted from simulated transient structures (Fig. S8).

**Acknowledgments:**

We thank Dr. John A. Jaszczak for providing single crystal bulk graphite samples, Dr. M. Harb and Dr. M. Mo for helpful discussion, and S. Yu for assistance in MD simulations. D. L. thanks the China Scholarships Council (CSC) Joint PhD Training Program for the financial support of studying abroad. This work was performed at the Center for Nanoscale Materials, a U.S. Department of Energy Office of Science User Facility, and supported by the U.S. Department of Energy, Office of Science, under Contract No. DE-AC02-06CH11357. The MeV-UED experiments were carried out at SLAC MeV-UED, U.S. Department of Energy Office of Science User Facilities, operated as part of the Linac Coherent Light Source at the SLAC National Accelerator Laboratory, supported by the U.S. Department of Energy, Office of Science, Office of Basic Energy Sciences under Contract No. DE-AC02-76SF00515.This work was also supported by the National Natural Science Foundation of China (NSFC, Grant No. 51771165, Grant No. 11805267), and the National Key R&D Program of China (YS2018YFA070119).